\documentclass[conference]{IEEEtran}
\usepackage{cite}
\usepackage[pdftex]{graphicx}
 \graphicspath{{./img}}
 \DeclareGraphicsExtensions{.pdf,.jpeg,.png,.eps}

\usepackage{color}
\usepackage{xspace}
\usepackage{amssymb}

\usepackage{textcomp}
\usepackage{hyperref}

\usepackage[cmex10]{amsmath}
\def\BibTeX{{\rm B\kern-.05em{\sc i\kern-.025em b}\kern-.08em
    T\kern-.1667em\lower.7ex\hbox{E}\kern-.125emX}}

\usepackage{algorithm, algpseudocode}

\usepackage{stfloats}

\newcommand{\ie}{\emph{i.e.}\xspace}
\newcommand{\eg}{\emph{e.g.}\xspace}
\newcommand{\etal}{et al.\xspace}

\newtheorem{theorem}{Theorem}

\newtheorem{definition}{Definition}
\newtheorem{remark}{Remark}

\newcommand{\netdelay}{\Delta}
\begin{document}
\title{Reducing Latency of DAG-based Consensus in the Asynchronous Setting via the UTXO Model}

\author{\IEEEauthorblockN{Keyang Liu}
\IEEEauthorblockA{The University of Tokyo\\
stephenkobylky2022@g.ecc.u-tokyo.ac.jp}
\and
\IEEEauthorblockN{Maxim Jourenko}
\IEEEauthorblockA{Tokyo Institute of Technology\\
jourenko.m.ab@m.titech.ac.jp}
\and
\IEEEauthorblockN{Mario Larangeira}
\IEEEauthorblockA{Tokyo Institute of Technology/IOG\\
mario@c.titech.ac.jp}}

\IEEEoverridecommandlockouts

\maketitle

\begin{abstract}
DAG-based consensus has attracted significant interest due to its high throughput in asynchronous network settings. However, existing protocols such as DAG-rider (Keidar et al., PODC 2021) and ``Narwhal and Tusk'' (Danezis et al., Eurosys 2022) face two undesired practical issues: (1) high transaction latency and (2) high cost to verify transaction outcomes. To address (1), this work introduces a novel commit rule based on the Unspent Transaction Output (UTXO) Data Model, which allows a node to predict the transaction results before triggering the commitment. We propose a new consensus algorithm named ``Board and Clerk'', which reduces the transaction latency by half for roughly 50\% of transactions according to our experiments. As the tolerance for faults escalates, more transactions can partake in this latency reduction. In addition, we also propose the Hyper-Block Model with two flexible proposing strategies to tackle (2): blocking and non-blocking. Using our proposed strategies, each node first predicts the transaction results if its proposal is committed and packs this result as a commitment in its proposal. The hyper-block packs the proposal's signature, and the consensus layer outputs to prove the transaction results.
\end{abstract}


\section{Introduction}

The appearance of blockchain has catalyzed the investigation of secure and resilient computation within decentralized systems fraught with Byzantine faults in both Nakamoto consensus and Byzantine Fault Tolerance (BFT) research. However, many public blockchains, such as Bitcoin~\cite{Nakamoto_2008}, rely on Proof of Work (PoW) and have been criticized due to energy consumption and constrained transaction throughput. In response to these challenges, alternative strategies offer a less computationally intensive mechanism. For example, executing the BFT algorithm based on a committee selected by Proof of Stack (PoS) has been adopted for balancing throughput and energy consumption. 

Numerous asynchronous BFT protocols have been studied in the past, and it constitutes a committee of nodes that are continuously proposing blocks (or proposals) and outputting a sequence of valid proposed blocks. Recently, Directed Acyclic Graph (DAG)-based consensus algorithms such as DAG-rider~\cite{Keidar_Kokoris-Kogias_Naor_Spiegelman_2021} and ``Narwhal and Tusk''~\cite{Danezis_Kokoris-Kogias_Sonnino_Spiegelman_2022} have demonstrated superior efficiency against existing solutions. However, two shortcomings persist: (1) There is high latency as one DAG consensus requires seven rounds of communication on average, and (2) The protocol's inability to discern failed transactions, which incurs additional costs (\eg additional communication for signatures from the committee) for practical application. This work offers a novel means to order transactions, thus improving BFT protocols' efficiency by relying on the UTXO Model.

\subsection{Related Works}

\paragraph{The BFT Revival}
The study of BFT consensus protocols has a long history~\cite{Castro_Liskov_2002, Yin_Malkhi_Reiter_Gueta_Abraham_2019, Duan_Zhang_2022, Sui_Duan_Zhang_2022}. These protocols are proved to tolerate $f$ byzantine faulty nodes among $N$ nodes, where $f<\frac{N}{3}$. Specifically, asynchronous BFT protocols~\cite{Kapron_Kempe_King_Saia_Sanwalani_2010,Miller_Xia_Croman_Shi_Song_2016,Lu_Lu_Tang_Wang_2020,Liu_Duan_Zhang_2020,Keidar_Kokoris-Kogias_Naor_Spiegelman_2021, Danezis_Kokoris-Kogias_Sonnino_Spiegelman_2022} usually requires all nodes to create proposals concurrently, thus avoid a single-node failure. 

The study of the BFT problem under asynchronous conditions started almost three decades ago with Ben-Or \etal~\cite{Ben-Or_Kelmer_Rabin_1994}. Since then, many solutions~\cite{Miller_Xia_Croman_Shi_Song_2016,Lu_Lu_Tang_Wang_2020,Zhang_Duan} have been proposed for solving this problem practically under the BFT framework. In 2021, Keidar~\etal~\cite{Keidar_Kokoris-Kogias_Naor_Spiegelman_2021} proposed DAG-rider, a novel approach that harnesses the DAG to organize all nodes' proposals and votes. This idea uses a growing sub-DAG to amortize the cost of making consensus, thereby drastically enhancing the speed of the asynchronous BFT algorithm. 

Building upon this foundation, Danezis~\etal~\cite{Danezis_Kokoris-Kogias_Sonnino_Spiegelman_2022} refines this design by merging the DAG with the gossip protocol to form the Narwhal Memory Pool, further dispersing the voting cost associated with the following proposal creation phase and catapults the system's throughput to a new peak (120K transactions per second on the Internet). Our work explores further improvement following these works.

\paragraph{The Blockchain UTXO Model}
Many Blockchain-based applications adopt the UTXO Model~\cite{Nakamoto_2008}, which is an alternative to the Account Model (\eg, a bank system and Ethereum~\cite{wood2014ethereum}). Chakravarty~\etal~\cite{Chakravarty_Chapman_2020} defined the \emph{Extended} UTXO Model as the stepping stone to provide smart contract capabilities to the Cardano Blockchain~\cite{EC:DGKR18,cardano}. Despite its wide use in the blockchain setting, to the best of our knowledge, not many protocols leverage the benefits of the UTXO Model to propose novel BFT protocol designs. This work showcases this approach by combining the BFT design and the \emph{Extended} UTXO Model to achieve better performance.

In the same spirit, Müller~\etal~\cite{MPPTSM_2022} proposed a UTXO-based ledger using the early cited inherent UTXO design parallelism to improve performance, thus speeding up the consensus process in the blockchain setting. Our work uses a similar strategy to shorten the latency but for the BFT setting. Although the UTXO model is widely used for Nakamoto consensus, to our best knowledge, this is the first work to propose its use in the BFT consensus algorithm design. 
\subsection{Our Contributions}

This work has two main contributions:

\begin{itemize}

    \item[1] We introduce a novel DAG-based commit rule designed explicitly for a UTXO-based transaction model. We propose a new consensus protocol named ``Board and Clerk" to replace Tusk, effectively reducing transaction latency. 
    
     \item[2] We introduce two new proposing flows and the Hyper Block Model designed to authenticate transaction results, which has zero communication overhead.
	
\end{itemize}

More concretely, the DAG-based consensus under asynchronous network conditions takes seven rounds on average to commit a sub-DAG. We show this latency is not mandatory for most transactions in our solution. Our work relies on the UTXO Model and proposes a new commitment rule for reducing transaction latency. Specifically, we equipped the DAG-based consensus and the Narwhal Memory Pool from~\cite{Danezis_Kokoris-Kogias_Sonnino_Spiegelman_2022} with a new commit rule for transactions under the UTXO Model. This new commit rule allows us to fast commit the transaction with predictable results (succeed or failed),  reducing the transaction confirmation latency without harming the correctness property. We propose Board to handle the fast commitment and change Tusk accordingly as Clerk. Our experiment shows that our Board and Clerk can achieve better latency than the original Tusk.

Furthermore, DAG-based consensus does not have communication overhead (\ie, there is no additional communication after the DAG is created), which means the order of transactions is unknown when nodes construct the DAG. This scenario may lead to the possible commitment of conflicting transactions (transactions that consume at least one same input, a.k.a. double spending). Since the consensus does not generate additional signatures after gossiping proposals, a client faces challenges in substantiating a transaction's success without all nodes' assistance. We propose two strategies to adjust the proposal flow, ensuring that all nodes can predict all transaction outcomes if the consensus round picks their proposal. Once the consensus output is confirmed, any honest node can generate proof to authenticate whether a transaction was successful according to the corresponding commitment.

\subsection{Organization}
Section~\ref{sec:preliminaries} provides more details about the problems we will resolve and the building blocks of our protocol. Sections~\ref{sec:txorder} and~\ref{sec:hyperblock} thoroughly describe our protocol. Finally, Section~\ref{sec:implementation} provides the analysis and experimental results, whereas our final remarks are in Section~\ref{sec:conclusion}.

\section{Preliminaries} \label{sec:preliminaries}
    
This section provides an overview of some fundamental concepts we utilize to describe our protocol. In particular, we revisit the asynchronous DAG-based BFT protocol design and the UTXO Model, which are the cornerstones of our proposal.

\subsection{Typical Asynchronous BFT Protocol Setting}

A BFT protocol typically proceeds in rounds involving multiple time slots and message exchanges. There are two types of participants: nodes and workers. Nodes participate in reaching distributed consensus by exchanging proposals with other nodes, while workers collect entries and deliver transaction batches to nodes. Each node has a reliable method for broadcasting messages~\cite{Bracha_Toueg_1985} and a uniform random coin for reaching consensus~\cite{Castro_Liskov_2002}. Furthermore, we assume the network is asynchronous, where messages between participants can take up to $\netdelay$ time slots to be received where $\netdelay$ is an unknown value. Each node knows the total number of participants and their identities, including their verification keys.

\subsection{DAG-based Asynchronous BFT Protocols}
\begin{figure}[!ht]
	\centering
	\includegraphics[width=3.5in]{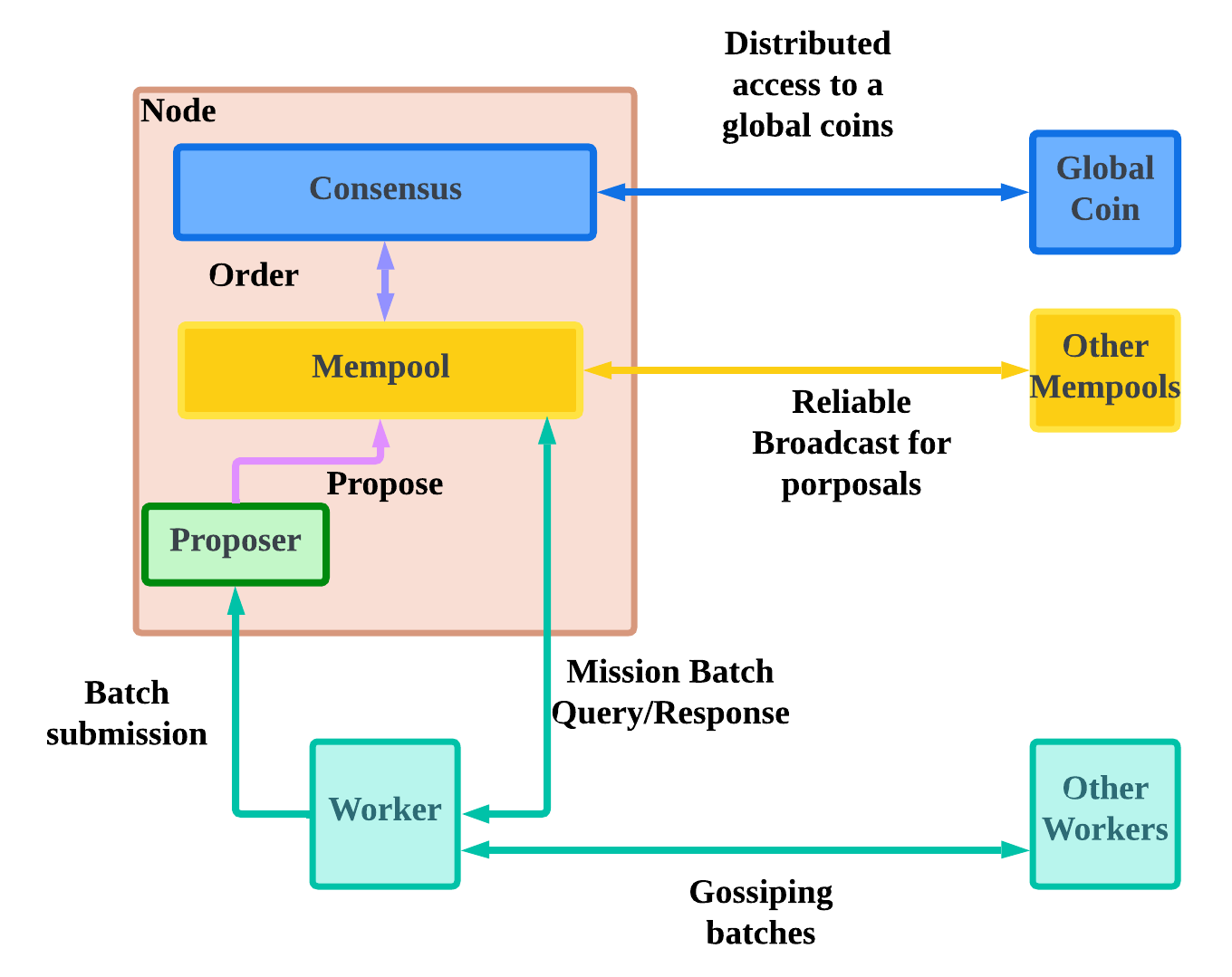}
	\caption{The communication model of a node in ``Narwhal and Tusk''~\cite{Danezis_Kokoris-Kogias_Sonnino_Spiegelman_2022}. The workers in green transferred the concrete transactions. The memory pool, in yellow, accomplishes the reliable broadcast. The consensus layer only needs access to the global coin for making decisions. }
	\label{fig_NT}
\end{figure}

These protocols are round-based distributed processes consisting of the DAG-making and decision-making phases, as in~\cite{Danezis_Kokoris-Kogias_Sonnino_Spiegelman_2022}. Each node receives a batch of transactions from workers and generates a proposal for consensus with other nodes. The architecture of a node is depicted in Figure~\ref{fig_NT}. In the later sections, we denote the Narwhal DAG as DAG.

More concretely, workers of a node will pack clients' transactions as batches and broadcast them to workers of other nodes using a Gossip Protocol, such as the one described in \cite{Bortnikov_2008}. Upon receiving enough responses from other workers, the worker sends the batch ID (i.e., hash values of underlying transactions) to its node. 

During the DAG-making phase of round $r$, the node aggregates the batch ID to create a proposal $P$ and $2f+1$ proposals' ID from round $r-1$. The node will use $P$ as input to initiate a Byzantine Reliable Broadcast (RBC)~\cite{LLRM_1982} instance. After this RBC instance is complete, the node stores $P$ in its Narwhal Memory Pool~\cite{Danezis_Kokoris-Kogias_Sonnino_Spiegelman_2022}. 

In the DAG-making phase, a node engages in at least $2f+1$ RBC instances and, therefore, stores $2f+1$ proposals. For proposals that contain unknown batches, the Narwhal asks the worker to synchronize the missing batches before continuing the RBC instance. The $2f+1$ stored proposals will be linked to its proposal in the next round. Thus, all proposals within Narwhal are stored as a round-based DAG, with an edge extending from one proposal to its $2f+1$ ancestors.

The DAG-based consensus algorithm chooses a sub-DAG from the Narwhal during the decision-making phase to fully utilize the results of the DAG-making phase. The Tusk consensus~\cite{Danezis_Kokoris-Kogias_Sonnino_Spiegelman_2022} randomly chooses one leader every two rounds from $N$ nodes, based on a universally random coin toss for being the candidate of a previous round's leader. If more than $f+1$ proposals link the selected candidate's proposal $L$, the protocol commits the sub-DAG linked by $L$. By leveraging a deterministic stable traversal algorithm, this $L$ uniquely determines the order of proposals within its sub-DAG, thereby dictating the output of the consensus algorithm.

Via the RBC instances, all honest nodes are assured of sharing an identical DAG eventually. If a leader's proposal from round $r$ has been referenced by $f+1$ proposals in round $r+1$, then the leader's proposal will always be in the sub-DAG of the leader of round $r+2$. Conversely, if a leader proposal receives less than $f+1$ followers, it may not be committed as a leader. This characteristic ensures the consistency of leader order and inherently verifies that the traversal of a DAG is deterministic. For a more detailed discussion, we refer the reader to the full work~\cite{Danezis_Kokoris-Kogias_Sonnino_Spiegelman_2022}.

\begin{figure}[!ht]
	\centering
	\includegraphics[width=3.5in]{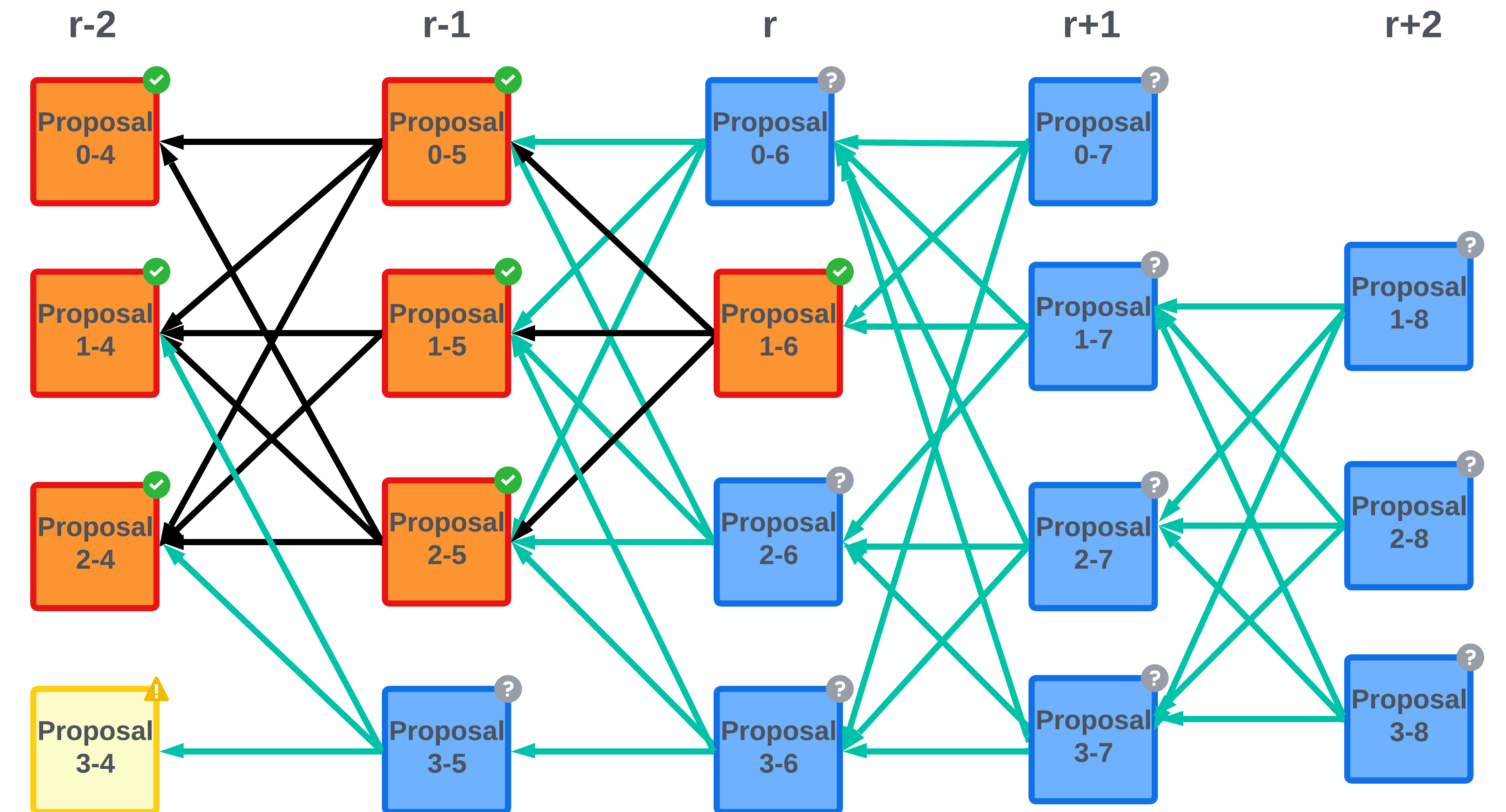}
	\caption{An example of proposal committing over DAG. The proposals are depicted as \#node-\#round, where \#node can be from $0$ to $3$ and round $r=6$.}
	\label{fig_DC}
\end{figure}

Figure~\ref{fig_DC} illustrates an example of DAG: the leader at round $r-2$ is the proposal 3-4, which does not have enough following blocks at round $r-1$. Thus, it cannot be committed as a leader (alert marked). The leader proposal at round $r$ is proposal 1-6, with two following proposals, which can commit all proposals in its sub-DAG (check-marked) through a deterministic traverse algorithm.
\ \\

 It is worth noticing that each node will have different views of the DAG at each round. For example, in Figure~\ref{fig_DC}, node one and node three have different views at round $r$ \ie, the sub-DAG that leads by proposals 1-6 and 3-6. We use the term ``observation'' to denote that a proposal or transaction is in such a local view, \ie, at round $r$, node 1 observes proposal 0-4 while node 3 does not.

\subsection{The UTXO Model}

The state of a UTXO-based ledger is represented by the set of UTXO that contain some data $\delta \in \{0,1\}^*$ as well as a script $s \in \{0,1\}^*$. A transaction $t$ consists of (1) a set of inputs that are references to UTXOs present on the ledger as well as witness data $w \in \{0,1\}^*$ and (2) a list of outputs that are newly defined UTXOs. Moreover, the ledger can define constraints on UTXO $\psi_u$ and constraints on transactions $\psi_t$~\cite{Chakravarty_Chapman_2020}. A transaction is considered \textit{internally valid} if (1) for each of its inputs $i$ holds that $s_i(w_i, \delta_i, t) \neq \bot$ where $w_i$ is the inputs witness data and $s_i, \delta_i$ are the referenced UTXOs script and data; (2) for each UTXO $u$ in its outputs holds that $\psi_u(u) \neq \bot$ and (3) that $\psi_t(t) \neq \bot$. A transaction is considered \textit{externally valid} if all UTXOs referenced by its inputs are in the ledger's current state. When a transaction is successfully committed to the ledger, all UTXOs referenced in its inputs are removed from the ledger's state, and all UTXOs in its outputs are added to the ledger's state. Two transactions are \textit{conflicting} if at least one UTXO is referenced in both transactions' inputs.

Verifying whether a transaction can be committed to the ledger relies only on verifying \textbf{(1) External validity} and \textbf{(2) Internal validity}~\cite{EC:DGKR18}. We note that internal validity does not depend on the ledger's global state and can be verified immediately upon transaction creation. Moreover, while external validity requires knowledge of the ledger's state, i.e., which UTXOs are present, it only relies on a small subset of the UTXOs present on the ledger. Thus, the external validity of a set of non-conflicting transactions can be computed in \textit{parallel}. This inherent parallelism of UTXO-based ledgers is an advantage to other ledger paradigms and is particularly interesting to this work. 

\subsection{Notation}
Table~\ref{tab_notation} lists the notations we used in this work:
\begin{table}[h!]
\centering
\caption{Table for notations}
\label{tab_notation}
\begin{tabular}{||c|p{0.7\linewidth}||} 
 \hline
 term & Description\\ [0.5ex] 
 \hline
 Tx & Short for a transaction, A digital exchange of value or information on the blockchain.\\
 \hline
 TXO & Short for transaction output, the minimum unit of information on the blockchain. It can only be created by one transaction and consumed by another transaction. \\
 \hline
 UTXO & Short for unspent transaction output, the TXO that is not consumed by other transactions.\\
 \hline
 Batch & A wrap of transactions for a consensus algorithm. \\
 \hline
 Proposal & A wrap of batches and other metadata that is proposed by a node for making consensus in the consensus algorithm.\\
 \hline
 Node & Process that participants of the consensus algorithm for the blockchain.\\
 \hline
 Worker & Process that collects transactions sent by clients and wraps them as a batch for nodes\\
 \hline
 Leader &  Leader is a pair of node IDs and rounds, the output of the DAG-based consensus. We abuse this notation for the corresponding proposal\\
 \hline
 $N$ & Number of nodes participants in the consensus algorithm \\ 
 \hline
 $f$ & number of maximum allowed byzantine nodes, usually satisfying $3f+1\geq N$ 
 \\
 \hline
\end{tabular}
\end{table}

\section{Board and Clerk}\label{sec:txorder}
   We first address the latency issue of existing DAG-based BFT consensus. According to \cite{Danezis_Kokoris-Kogias_Sonnino_Spiegelman_2022}, the average latency of Tusk under an asynchronous setting is seven asynchronous rounds. Our proposal uses the UTXO model to finalize some transactions' results before this number and thus improve the user experience. Before describing our protocol, we present an initial intuition regarding the interplay between the DAG data structure and the UTXO Model.

\subsection{Intuition: DAG meets UTXO}

Our first proposal is a new transaction commit rule that compares the ordering decided by the random leader in Tusk. The UTXO Model is designed to support parallelism, so we do not need a strict sequential order of all transactions. Considering that workers can ensure internal validity beforehand, the consensus algorithm only needs to identify double-spending transactions and agree on which transactions are invalid. 

Currently, the existing DAG-based consensus algorithms rely on deterministic DAG traversal algorithms to determine the transaction order and the external validity. However, the proposer of the leader proposal often decides the order almost randomly based on how it arranged the proposals from the previous round. This approach requires significant work to determine the transaction's external validity. 

Therefore, we propose a new commit rule that does not rely on the order of transactions in the batches, making the results of transactions more predictable before their commitment. We use the parallelism of the UTXO model in our approach. Given that the proposed phase has checked internal validity, our commit rule uses the DAG to finalize the external validity independent of the consensus algorithm. Consequently, any node can be confident about the results of most transactions (except double-spending transactions) after a certain number of rounds, which enables them to process the following transactions and, ultimately, reduce the actual latency of transaction commitment. 

\subsection{Transaction States}

We use the transaction state to represent its lifecycle in memory. The states can be:

\begin{itemize}
    \item \textbf{Verified}: A received transaction that has passed the internal validity check by a worker;
    
    \item \textbf{Submitted}: A verified transaction included in at least one single proposal;
    
    \item \textbf{Fast-committed}:  A submitted transaction that will be committed in the future with a confirmed result;
    
    \item \textbf{Committed}: A submitted transaction that the consensus algorithm has committed.
\end{itemize}

We use the fast-committed state to process transactions and apply changes to the state before committing them via the consensus algorithm. The fast-committed transactions use a shorter internal path to execute and enable the node to verify its following transactions. This feature is especially advantageous when the consensus algorithm takes a long time to select appropriate leaders for committing the DAG. 

\subsection{The Novel Commit Rule}

A local node must determine whether a transaction will succeed or fail to enable a fast-committed channel. Therefore, a procedure to decide the external validity of a transaction is required to reduce uncertainty when extending the DAG. Hence, we have made the following two changes to the proposal flow compared to Narwhal and Tusk:
\begin{enumerate}
    \item We require that any valid batch of transactions does not contain conflicting transactions. A batch is a wrapper of transactions that are submitted through a worker. It is reasonable to assume that the worker has verified all transactions in the batch, ensuring that transactions spending the same UTXO are not kept. It should be noted that different batches may have conflicting transactions since they may come from different workers.

    \item We require that a node's proposal for round $r$ always includes its proposal for round $r-1$. This requirement ensures the basic consistency of the node's proposal. If a node's proposal at round $r$ is committed, so is its proposal at round $r-1$. This significantly improves the reliability of the projection of the external validity decision.
\end{enumerate}
Now, we provide some definitions describing our approach to determine external validity.

\begin{definition}
For a sub-DAG that begins with a node's proposal at round $r$ containing a transaction $tx$ for the first time, we say that the node votes for $tx$ at round $r$.
\end{definition}

\begin{figure}[htbp]
	\includegraphics[width=\linewidth]{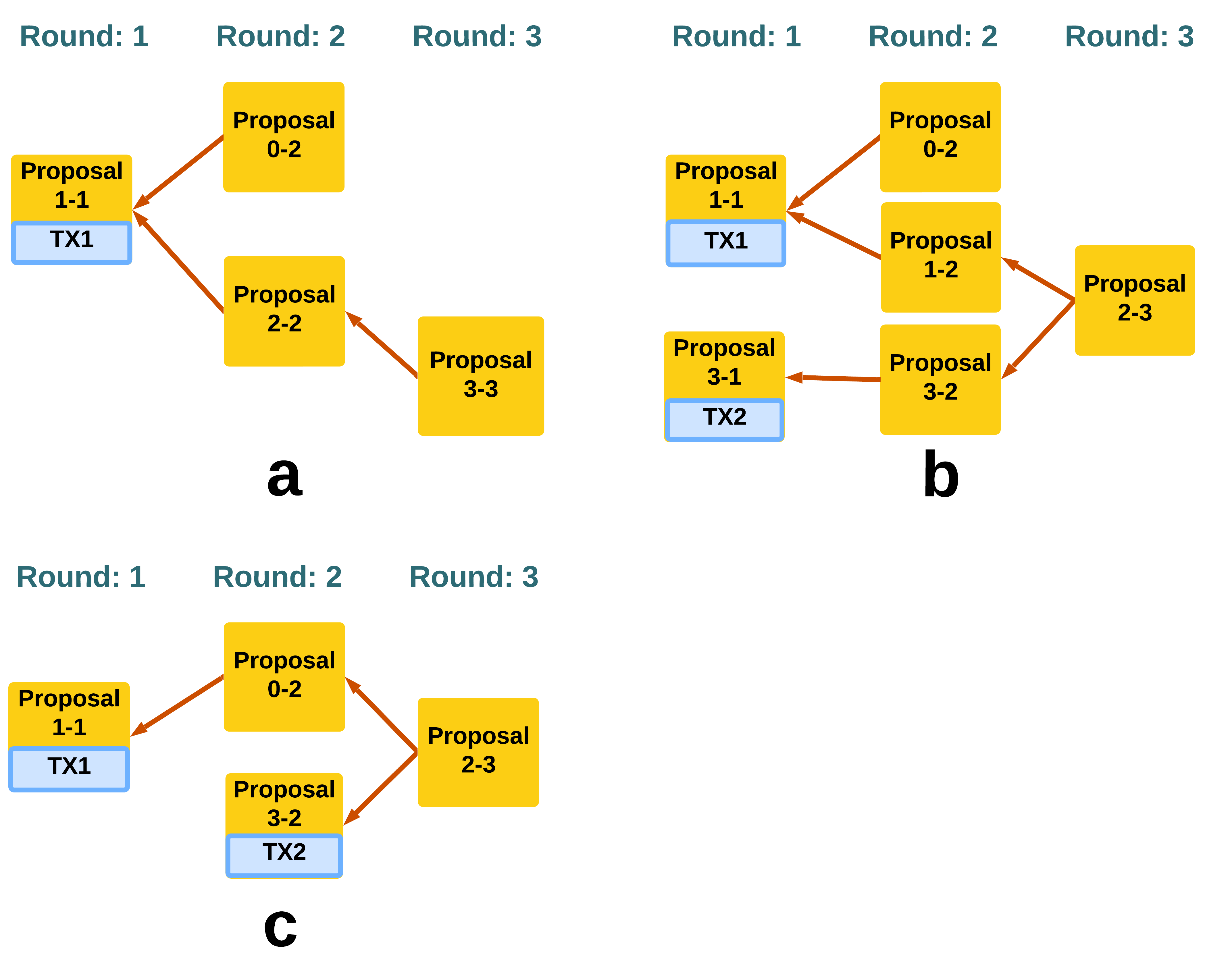}
	\centering
	\caption{Three examples for transaction votes. (a) Node 1 proposes $tx_1$ at round 1. Node 0 and Node 2 vote it in round 2, and Node 3 votes in round 3; (b) Node 3 proposes for $tx_2$ in round 1, and Node 2 votes for both transactions in round 3; (c) Node 2 still votes for $tx_1$ and $tx_2$ at round 3. }
	\label{fig_TX}
\end{figure}

A node generally votes for a transaction in one of two cases: (1) its proposal includes the transaction, or (2) its proposal links to a proposal from another node that already voted for the transaction. The vote is only kept for the minimum round. To be noticed, a node may vote for conflicting transactions at the same round; in this case, its vote is counted for both transactions. Figure~\ref{fig_TX} shows some typical cases for voting. 

\begin{definition}~\label{def:ord}
For any set of transactions, we define an ordering relationship $Ord$ such that there is only one stable sort of the set under $Ord$.
\label{def_tx_ord}
\end{definition}
A straightforward example is ordering according to the hexadecimal representation of transaction IDs.

\begin{definition}[External Validity-Commit Rule]
For all transactions inside a sub-DAG to be committed by the consensus algorithm at even rounds, a transaction is successful only if more nodes have voted for it before any other transactions conflict. In other words, let $tx.I$ be the inputs of $tx$ and $\mathsf{TX_S}$ be the set of all submitted transactions. We commit $tx$ if:

$$\forall tx' \in \mathsf{TX_S} \; : \; tx'.I \cap tx.I = \varnothing.$$
Furthermore, it is also committed if
\begin{equation}
    |\{\text{nodes vote tx first}\}| > |\{\text{nodes vote }tx'\text{ first}\}|,\nonumber
\end{equation}
or otherwise if
\begin{equation}
\begin{split}
  & tx >_{Ord} tx', \\
  & |\{\text{nodes vote tx first}\}| = |\{\text{nodes vote tx' first}\}|.\nonumber \\
\end{split}
\end{equation}
\label{def_tx_rule}
\end{definition}

Definition~\ref{def_tx_rule} allows for a deterministic outcome of the transaction's external validity. Transactions without conflicts are committed directly. For transactions with conflicts, we commit according to the votes from nodes, and any tie will be resolved by the deterministic order $Ord$ (given by Definition~\ref{def:ord}).

Compared to~\cite{Danezis_Kokoris-Kogias_Sonnino_Spiegelman_2022}, we only change the commit rule. This new commit rule enables the node to pre-compute the votes for each transaction before the consensus decides the leader. We present Theorem~\ref{The_SuccessCase} in the center of our proposal correctness. Here, a node observes a transaction when a proposal that contains that transaction has been added to its DAG.
\begin{theorem}
    \label{The_SuccessCase}
    When a node added a proposal $p$ to its DAG at an odd round $r$, if a transaction $tx$ has more than $2f+1$ votes from nodes and no valid conflicting transactions were observed, then $tx$ will eventually be committed as a successful transaction according to Definition~\ref{def_tx_rule}.
\end{theorem}

\begin{IEEEproof}
    Recall that Definition~\ref{def_tx_rule} states that a transaction that has been included in more nodes earlier than other conflicting transactions will be committed, and the consensus layer commits a leader proposal only if $f+1$ proposals have included it in the next odd round.

    Firstly, we demonstrate that the transaction will eventually be committed. As there are at least $2f+1$ nodes whose proposals indicate they have voted for the transaction $tx$, any proposal at round $r+1$ will contain the transaction. Thus, $tx$ will be committed by some proposal eventually.
    
    Next, we show this transaction will be committed successfully. If any proposal is to be committed before round $r$, it must have $f+1$ subsequent proposals, and thus, it will be in the sub-DAG starting from $p$. Hence, the node will observe any valid conflicting transaction that can be committed before round $r$. If a proposal $p'$ after round $r$ is selected as a leader, then its sub-DAG will have at least $2f+1$ proposals from round $r$, which implies that at most $f$ nodes will vote for conflicting transactions while at least $f+1$ will vote for $tx$. According to our commit rule, $p'$ can only commit $tx$.
\end{IEEEproof}

A proposal in an even round cannot trigger fast-commit since its vote on that round may not impact the leader's choice in the same round. The leader may encounter a tied vote in its sub-DAG case and commit a different transaction. According to Theorem~\ref{The_SuccessCase}, we can define the fast commit rule.

\begin{definition}[Fast Commit Rule]
    A transaction is fast committed with a successful result at an odd round if $2f+1$ nodes have voted it and no valid conflicting transactions are observed.
    \label{def_fast_commit}
\end{definition}

This fast commit rule is derived from our new commit rule and Theorem~\ref{The_SuccessCase}, which does not affect the security claim of Tusk. Estimating the expected latency reduction using this fast commit rule is non-trivial. According to our experiments in section~\ref{sec:implementation}, we can reduce about half of the transactions' latency by 50\%, i.e., an average 25\% latency reduction.

\subsection{Our Construction: Board and Clerk} 

We propose a new consensus algorithm called ``Board and Clerk" to replace Tusk for applying our new commit rule. At a high level, Board decides on transaction results and fast-commit transactions, and Clerk agrees on the formal commitment. The algorithm operates as follows: After Narwhal commits a proposal, Board counts the votes from that proposal, keeps it in storage, and fast commits transactions if possible. Once the algorithm reaches the commit rule (\ie when processing a proposal of even round $r>4$), Clerk uses a global coin to select the leader. If the leader is valid according to the DAG, Clerk will output all uncommitted proposals from the leader's sub-dag, and Board can compute the transaction result accordingly for commitment.

Board stores all the vote results and conflict information. More specifically, Board stores a map of \{transaction Id: vote record\} and a map of \{batch: vote record\}. Each vote record is a hash map between \{node Id: round\} that records the round number that a node voted for the element (transaction or batch). We use Board.AddTxVotes and Board.AddBatchVotes functions to denote updating the vote (insert the round for the node ID). 

Board provides two open APIs: first, when a new proposal is added to the DAG, Board counts votes for the author node. Since proposals are added to the DAG after its ancestors, the first vote from a node to an element is always smaller than the later votes from that node to the same element. Algorithm~\ref{Alg_Count_Vote} highlights the flow for counting transactions' votes, which triggers the fast commitment channel.

\begin{algorithm}[htbp]
	\caption{Algorithm for counting transaction votes}
	\label{Alg_Count_Vote}
	\begin{algorithmic}[1]
		\Require Proposal P
        
        \Function{Board.Process}{P}
        
        \State Initiate list B
        \State P.round $\rightarrow$ round
        \State P.author $\rightarrow$ voter
        \For{ batch in P.batches}
                \If{!Board.HaveSeen(batch)} 
                    \State add the batch in B
                \EndIf
        \EndFor

        \For{ batch in B}  
            \For{ tx in batch}
                \State Board.Add(tx.txos, tx)
                \If{Any TXO has more than one tx}
                    \State Add tx to conflicted set
                \EndIf
            \EndFor
            \State Board.AddTxVotes(batch.txs, voter, round)
        \EndFor

        \Comment{Counting Votes for sub-dag proposal}
        \State P.parents $\rightarrow$ C
        
        \While{C is not empty} 
            \State C.Pop() $\rightarrow$ parent
            \State Board.AddBatchVotes(parent.batches, voter, round)
            \If{any vote changed}
                \Comment{new batch}
                \State C.Concat(parent.parents)
                \State Board.AddTxVotes(batch.txs, voter, round)
            \EndIf
        \EndWhile

        \If{round is odd}
            \Comment{Fast Commit}
            \State Filter out transactions that are not conflicted and have more than 2f+1 votes.
            \State Fast commit those transactions
        \EndIf
        \EndFunction

\end{algorithmic}
\end{algorithm}

To aid the decision process of transaction states, we utilize internal states. To achieve this, we require the storage of node votes for different batches and transactions, with the added benefit of serving as a shortcut to avoid duplicated batches. Additionally, we must maintain records of conflicting transactions and the existing TXO-TX pairs, thus preventing the missing of any vital information regarding conflicts. To ensure smooth operation, we also use a cache to store transactions fast-committed to reject later conflicting transactions.

The second API for the Board is triggered when Clerk commits a sub-DAG by the consensus, resulting in Board committing all transactions in the sub-DAG as required. Since the consensus algorithm only decides the leader for the previous round, we cannot directly use the latest vote counts, which include votes that are not to be committed. To avoid recounting votes, we define a frontier for each proposal.

\begin{definition}
A frontier $F$ of a proposal $p$ is a map between node IDs and their latest proposal round in the sub-DAG that begins with $p$.
\end{definition}

We can remove the vote records after the corresponding round in the frontier since the current commitment will not include those vote records. The flow for counting transaction results is outlined in Algorithm~\ref{Alg_TX_Result}.

\begin{algorithm*}[htbp]
	\caption{Algorithm for getting transaction results of a sub-DAG}
	\label{Alg_TX_Result}
	\begin{algorithmic}[1]
		\Require A list of proposals P
            \Ensure A set of successful transactions ST and a set of failed transactions FT
            
        \Function{Board.Commit}{P}
        \State Init empty ST, FT
        \State Set T as a list of all uncommitted TXs in the sub-DAG of P 
        
        \State Board.ConflictedTX() $\longrightarrow$ CT
        \Comment{CT is the dictionary of conflicted TXO and its conflicted transaction pair according to TXO-TXs}
        \State T.IntersectWith(CT.values) $\longrightarrow$ TxToProcess
        \If{ TxToProcess = $\phi$}
            \Comment{We can directly commit all as success}
            \State ST.Add(T)
            \State Return ST, FT
        \Else
            \State ST.Add(T.Except(TxToProcess))
        \EndIf

        \State Set M as a map of $<$TXO,TX$>$ with values from TxToProcess
        \State M.sortbyKey()
        \Comment{use a stable sort}
        \State Compute frontier $F(P)$
        \For{(txo,txs) in M}
            \State Prune transaction vote records by F and remove vote records from nodes that are not earliest.
            \State Sort txs with vote counts and $Ord$
            \State Push the first tx in txs to ST and the remaining to FT
            \For{ (txo', txs') in M}
            \Comment{We need to check if other conflicted sets have been affected by this result}
                \If{tx contains txo'}
                    \State Push other tx' in txs' to FT
                \EndIf
                \State txs' = txs'.disjoint(txs)
            \EndFor
        \EndFor

        \State Fast Commit other transactions according to ST, FT
        \Comment{Some transactions might meet the fast commit rule by now, (\eg, the conflicted $tx$ is failed) or it is not processed now but doomed to fail (\eg, the conflicted $tx$ is succeeded now)}
        \State Board.CleanUp
        
 	\State Return ST, FT
  \EndFunction

\end{algorithmic}
 
\end{algorithm*}

\subsection{Theoretical Analysis}

This work inherits the adversary model from Tusk~\cite{Danezis_Kokoris-Kogias_Sonnino_Spiegelman_2022}. Given $n$ parties in an asynchronous network of our consensus algorithm, a computationally bounded adversary can corrupt up to $f<n/3$ parties, denoted as faulty nodes. Our protocol shares the same rule for committing leader block as Tusk~\cite{Danezis_Kokoris-Kogias_Sonnino_Spiegelman_2022}, which allows our protocol to have the same security and liveness claim. The transaction result is determined by Algorithm~\ref{Alg_TX_Result}, which is deterministic according to the sequence of the leader block. Therefore, our protocol is as secure and live as Tusk~\cite{Danezis_Kokoris-Kogias_Sonnino_Spiegelman_2022}. 

Compared to Tusk, our protocol introduces additional computation and storage costs to reduce the latency of fast-committed transactions. Generally, Board needs an additional space linear to the number of transactions for storing votes. Algorithm~\ref{Alg_Count_Vote} requires vote updating for each batch and transaction, which is linear to the number of nodes $N$ given the transaction number is also linear to the number of proposers. Algorithm~\ref{Alg_TX_Result} introduces a computation overhead of $O(m^2)$ for $m$ conflicted transactions. This overhead is negligible since $m$ is relatively small compared to the total transaction number. Some additional trade-off between computation and storage is possible. For example, in Algorithm~\ref{Alg_Count_Vote}, we use additional batch votes to filter out transactions with new votes. When committing proposals. Nodes running our protocol need more memory and computation power than Tusk. In the experiment section (Figure~\ref{fig_scale_nodes}), we will see the formal block latency of our protocol has little difference with Tusk if the computation and memory are not the bottlenecks.

\section{Hyper-Block Model}\label{sec:hyperblock}
   Our second proposal is a Hyper-Block Model that allows for generating proofs for transaction results. Our solution adheres to the architecture demonstrated in Figure~\ref{fig_NT} and integrates the fast-commit rules from Definition~\ref{def_fast_commit}. Previous DAG-based consensus protocols focus on creating consensus by ordering transactions such that proof of whether transactions were included in the DAG and which conflicting transactions succeed or fail is available only after a DAG is formed. The purpose of the Hyper-Block model is to create such proof for the status of a transaction in parallel to form the DAG, thus reducing the latency between the commitment and the inclusion of a transaction in the consensus. Note that our approach does not produce any communication overhead such that its impact on the performance of DAG creation is minimal.

At a high level, a hyper-block consists of three parts: 1) proof of leadership, 2) a commitment to batches in the sub-DAG, and 3) a commitment to the failed transaction set. The consortium will sign all contents to enable any client to use the hyper-block and verify whether its transaction is correct. The commitment of the set can be any proper cryptographic aggregator that satisfies the requirements. Specifically, each proof of transaction result contains three parts. 
\begin{itemize} 
    \item \textbf{Block membership}: Proof that the transaction is in a batch contained by a hyper-block; 
    
    \item \textbf{Result Proof}: A demonstration of whether the transaction is a member or non-member of the corresponding failed transaction set in the hyper-block; 
    
    \item \textbf{Leadership Proof}: Proof of leadership for the protocol. 
\end{itemize}

The primary technical challenge arises from preparing the hyper-block without disrupting the consensus protocol, which only decides the leader proposal index. To overcome this problem, we need to accomplish two tasks: Firstly, we need to obtain the transaction results to commit the failed transactions. Secondly, we must ensure that byzantine nodes cannot forge a hyper-block that validates a failed transaction.

We propose two strategies to achieve this: 1) A blocking strategy that synchronizes the DAG and consensus layers to execute a process similar to a view-change in other BFT protocols, and 2) a non-blocking strategy that commits all possible and valid output paths through the consensus algorithm.

Here we list some sub-functions we use in our blocking and non-blocking algorithms, along with implementation remarks: 
\begin{itemize}
    \item MakeProposal: Take a round $r$ as inputs and return a proposal $p$ that patches at least $2f+1$ proposals from the previous round and a bundle of batches received from its workers; 
    
    \item TakeLeader: Take a round $r$ as inputs, await the consensus layer output the leader of round $r$. If the leader is not in the DAG, recursively outputs the latest leader in the DAG;
    
    \item SimulateCommit: Take a proposal and the previous leader as input. It outputs a list of committed batches and a set of failed transaction IDs. The flow is almost the same as Algorithm~\ref{Alg_TX_Result} without side effects, thus should be implemented as a function of Board;
    
    \item CommitList: Take a list as input and return one value as output commitment. It can be implemented as a cryptographic accumulator that supports both membership and non-membership proof;

    \item AttachWitness: Including some witnesses to a proposal $p$ for acquiring signatures.
\end{itemize}

Now, we thoroughly detail both strategies.

\subsection{The Blocking Strategy}
The blocking strategy modifies the clerk consensus algorithm and proposes rules. For the consensus algorithm, if the first proposal of an odd round of $r\geq 5$ is added to the DAG, the consensus layer will automatically trigger the protocol to select the leader for the round $r-1$. 

Secondly, when the proposer initiates a proposal for any even round of $r+1 > 4$, it will first collect batches and wrap them properly. Next, the proposer will block the proposal and await the selection of a leader from round $r-1$. After the proposer becomes aware of the selection of the round leader, it will simulate the transaction results of its proposal as if this proposal is committed to the future leader. This simulation will compute the corresponding failed transaction commitment and batch set commitment. Subsequently, the two commitments are wrapped together with the proposal and dispatched to the broadcast phase. The flow for the proposing process is presented in Algorithm~\ref{Alg_Proposing}.

 \begin{algorithm}[htbp]
 	\caption{Blocking algorithm for proposing}
 	\label{Alg_Proposing}
 	\begin{algorithmic}[1]
 		\Require Round $r$
 		\Ensure Proposal $p$ for round $r$
 		
            \Function{Proposer.Proposing}{$r$}
 		\State MakeProposal($r$) $\longrightarrow p$
            \State TakeLeader($r-2$).await() $\longrightarrow L$
            \State Board.SimulateCommit($p, L$) $\longrightarrow (B,F)$
 		\State CommitList($F$) $\longrightarrow wf$
            \State CommitList($B$) $\longrightarrow wb$
 		\State AttachWitness($p$,$wb,wf$)
 		\State Return $p$
            \EndFunction
 	\end{algorithmic}
 \end{algorithm}

Upon receiving the proposal from the other nodes, the initial action is to verify the DAG history and confirm the correct calculation of the failed transaction commitment. Confirmation is done by signing the proposal ID, the commitment of failed transactions, and the commitment of batch set. The signed information is used to generate the certificate of the proposal in the DAG. Once the proposal is successfully committed to a leader, each node independently assembles the hyperblock by fetching this signature and the output of the consensus layer.

Our blocking strategy does not affect the security claims of the Tusk algorithm, as the election process is independent.  

\begin{remark}
The counting process can be accelerated for efficiency by utilizing dynamic programming methods to cache the votes for conflicted transactions (M in line 13 of Algorithm~\ref{Alg_TX_Result}). The cached outcome can later be removed after the commitment of corresponding leaders.
\end{remark}

\subsection{The Non-Blocking Strategy}

In an asynchronous network setting, waiting for the consensus layer results in the blocking strategy may not be tolerable. Therefore, we propose a non-blocking strategy to minimize the potential delays due to the network.

We use a commit and prove flow technique to create the hyperblock in the non-blocking strategy. The proposer maintains a leader DAG by the present DAG. The difference is that the leader DAG comprises proposals only for even rounds (\ie, rounds for selecting leaders). Every two proposals are connected only if they are connected in the DAG. Moreover, whenever a leader is chosen for a round $r$, the proposals other than the selected one are removed from the leader DAG instantaneously. The leader DAG can be seen as comprising all the probable leader proposals of the current DAG.

In the leader DAG, we store batch and failed transaction commitments. Each proposal has a vector for these two commitments; every position in the vector reflects a path in the leader DAG from the current proposal to the previously committed leader (if the leader is not in the DAG), given that the corresponding proposals on the path are chosen as the leaders for the corresponding rounds. 

\begin{figure}[htbp]
	\centering
	\includegraphics[width=\linewidth]{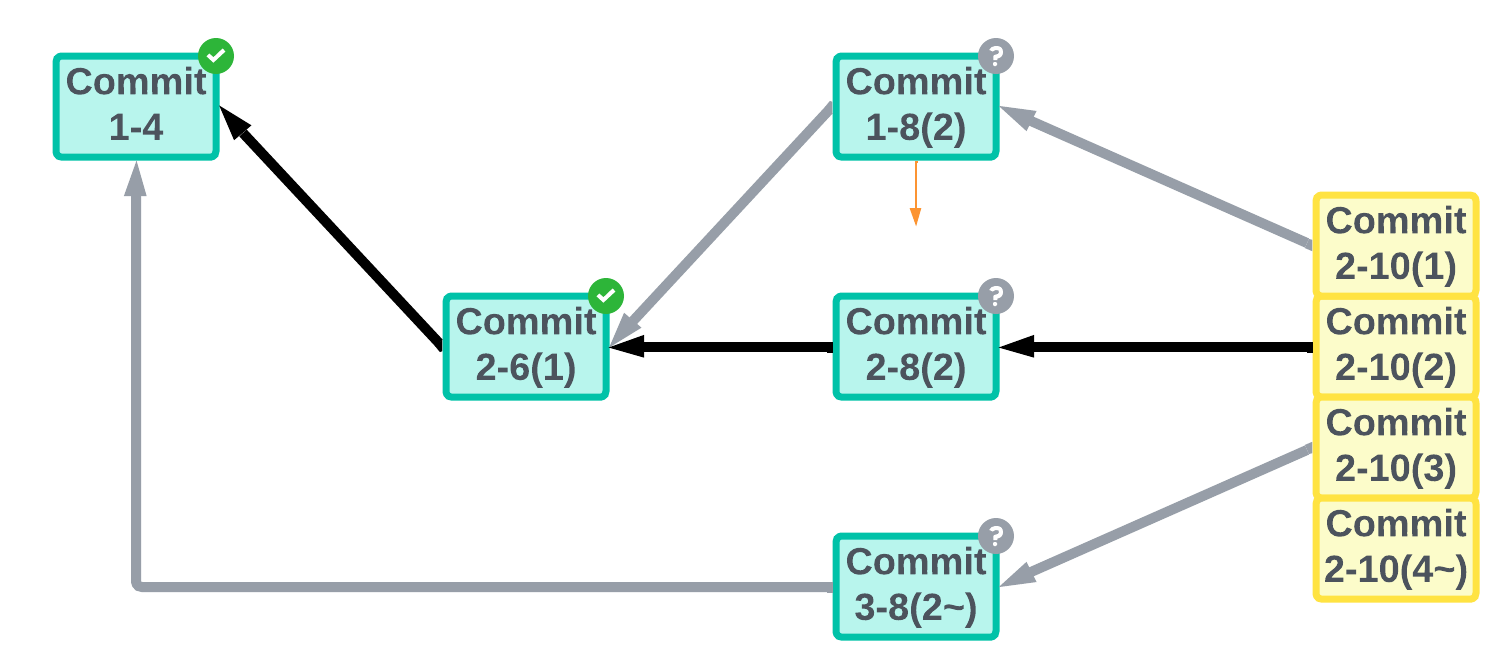}
	\caption{An example of a leader DAG. The dark line represents the path of final commitment, and the gray lines represent the candidates' paths during proposing.}
	\label{fig_leader_dag}
\end{figure}

Figure~\ref{fig_leader_dag} shows an instance of a leader DAG and its verification path based on the consensus layer results. When proposing for round 10, we assume the leaders of round 4,6 are known(checked), and the leader of round 8 is unknown (question marked). The new proposal at round 10 has four possible cases for the round leader at round 8. It will create corresponding commits and form them as a vector for proposing. For proposals at round 8, which already know the leader of round 6 (checked), Node 2 will contain a commitment for it and be used by the proposer in round 10. On the other hand, Node 3 has not yet seen proposals 2-6, so its commitment to the older leader 1-4 (checked) will be used for proposing. If the 2-8 is the eventual new leader, the black line represents the commitments that all users eventually use.

We require that a leader DAG supports the following functions: First, Candidates($r$) returns a list of candidates leaders for round $r$. If the leader is not selected, it should be all proposals in round $r$ plus Candidate($r-2$) in a recursive way. Second, Add($p, V$) adds the proposal $p$ to the leader DAG with a vector $V$ of commitment of batches and transactions. In addition, the leader DAG should also support some clean-up functions to remove non-leader or committed proposals accordingly.

The proposal contains the vector created according to the paths of the leader DAG. After receiving a proposal, a node verifies the vector and then commits it as a Merkle tree, sending the root's signature as evidence back to the proposer. This evidence, along with the consensus layer output, will eventually be used to verify the correctness of the batch commitment and the failed transactions commitment made by the node if this proposal is the leading one in the future. Algorithm~\ref{Alg_Proposing_NB} outlines the proposal procedure in the non-blocking strategy.

 \begin{algorithm}[htb]
 	\caption{Non-Blocking algorithm for proposing}
 	\label{Alg_Proposing_NB}
 	\begin{algorithmic}[1]
 		\Require Round $r$, Leader DAG $LD$
 		\Ensure Proposal $p$ for round $r$
 		\Function{Proposer.Proposing}{$r,LD$}
 		\State MakeProposal($r$) $\longrightarrow p$
            \State Initiate a vector $V$
            \For{ $L$ in $LD$.candidates($r-2$)}
                \State Board.SimulateCommit($p,L$) $\longrightarrow (B,F)$
     		\State CommitList($F$) $\longrightarrow wf$
                \State CommitList($B$) $\longrightarrow wb$
                \State $V$.Push($wf,wb$)
            \EndFor
            \State $LD$.Add($p,V$)
 		\State AttachWitness($p$,$V$) 
 		\State Return $p$
        \EndFunction
 	\end{algorithmic}
 \end{algorithm}

\subsection{Theoretical Analysis}

Both strategies only add one commitment for successful and failed transactions to the proposal. Narwhal will collect sufficient signatures for the proposal, including this commitment, so nodes can prove the result of arbitrary transactions through this commitment(\eg, proving a transaction has a valid open to the successful or failed set). In the non-blocking strategy, the prover must provide additional proof of the previous leader since it commits all possible results for different leaders. 

Since both strategies are deterministic processes, each proposal's communication overhead is just $O(1)$. Algorithm~\ref{Alg_TX_Result} dominates the computational complexity of both the blocking and non-blocking strategies. For the blocking strategy, it should be noted that this process triggers at least $2f+1$ simulations during the RBCs phases of each round. However, the result can be cached for the formal commit process. Unlike the blocking strategy, the non-blocking strategy involves a polynomial scale of possible paths to commit. However, in practice, if the leader of round $r-4$ is evident by round $r$, the proposer needs to calculate at most $2f+2$ possibilities for its vector, given that each proposal's commitment at round $r-2$ is already confirmed. Thus, the total computation cost for each leader round is at most $O(n^2)$ in this case. 

\begin{remark}
    Practically, we can hybrid non-blocking and blocking strategies. We can safely choose the non-blocking strategy when there are no or limited conflict transactions. Once the conflicted transaction increases, a blocking strategy can cool down the proposing phase and reduce the computation overhead.
\end{remark}

\section{Implementation and Evaluation}\label{sec:implementation}
    In this section, we estimate the latency reduction of our Board and Clerk scheme and compare it with existing protocols. The Hyper-Block Model is omitted since it is a pure gain of function that relies on heavier local computation. The main target of our experiment is to demonstrate how fast commit rule can affect the latency under different \textbf{TPS}, \textbf{number of nodes}, and \textbf{number of faulty nodes}. Since our scheme does not introduce communication overhead, we can estimate all these factors in a cheaper local testbed experiment. We have implemented Board and Clerk based on the code base of MystenLabs\cite{narwhal_code}. 

Our implementation solely modifies the local execution flow without introducing additional communication costs. Thus, we only perform local tests to evaluate the latency reduction achieved by our algorithm. Our Clerk algorithm follows the original Tusk to commit the block~\cite{Danezis_Kokoris-Kogias_Sonnino_Spiegelman_2022}. To compare the results, we analyze the performance of our algorithm against Tusk's and Bullshark's~\cite{Spiegelman_Giridharan_Sonnino_2022}. Bullshark is a variant of Tusk optimized for the partial synchronous setting.

\begin{figure}[!ht]
	\centering
	\includegraphics[width=\linewidth]{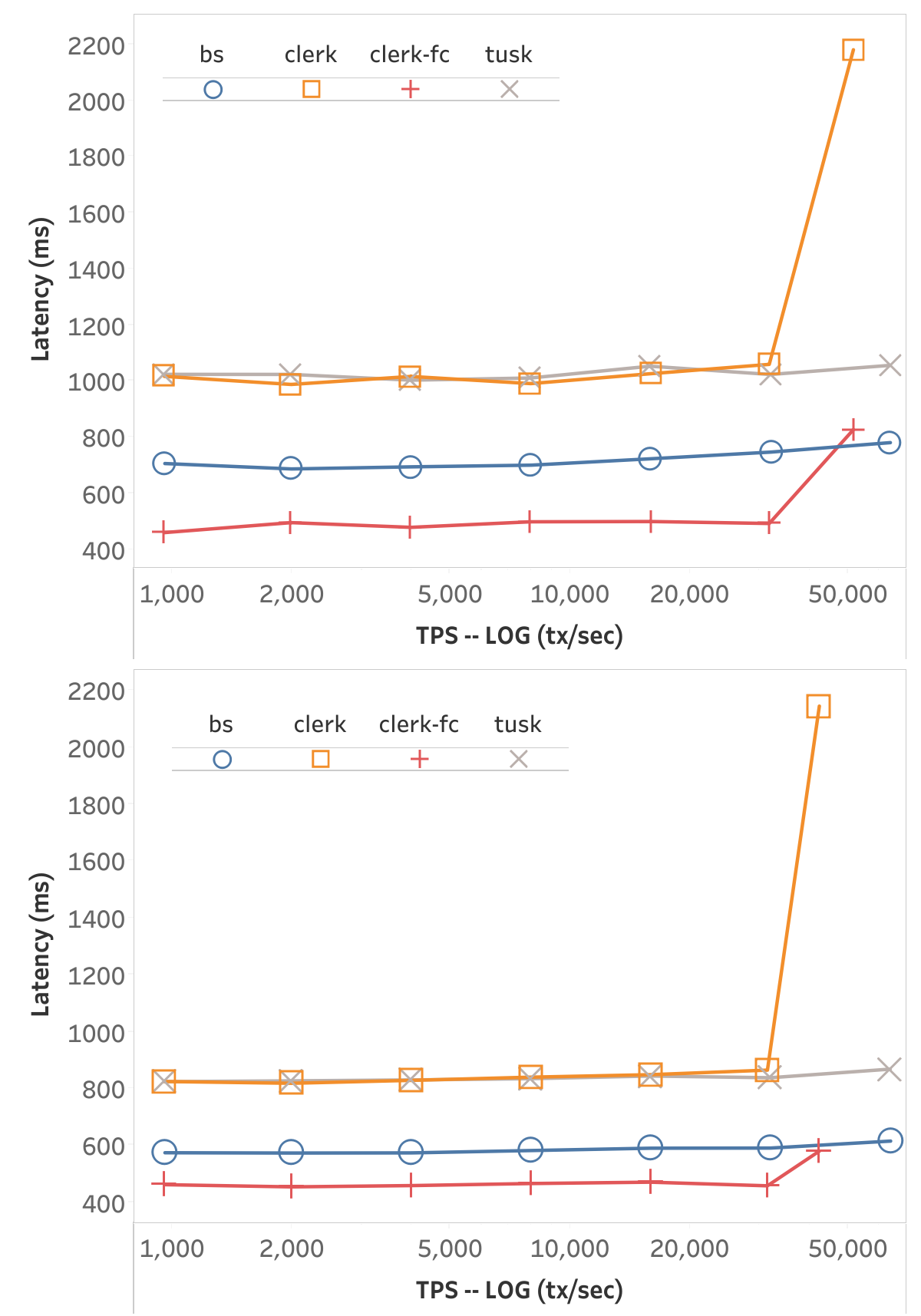}
	\caption{TPS-Latency graph (top: asynchronous setting, bottom: partial synchronous setting).  
    }
	\label{fig_latency_tps}
\end{figure}

The simulation was conducted on a desktop machine with an Intel i7-13700K CPU, 32GB DDR5 memory, and 2TB SSD storage. We test different TPS for a 4-node system, and Figure~\ref{fig_latency_tps} illustrates our latency reduction in both the asynchronous setting (top), where $\netdelay$ is unknown, and the partial-synchronous setting (bottom), where it is known and used as the timeout parameter. In Figure~\ref{fig_latency_tps}, the latency is measured by the end-to-end time between the client's submission of transactions and the node processing them and updating the state. The clerk (block marked) line shows the latency of transactions that are not fast-committed. The clerk-fc (plus marked) line shows the latency of fast-committed transactions.

We draw two observations from this result: First, our algorithm keeps the regular commitment time compared to the original Tusk protocol. In addition, our fast commitment channel \textbf{reduces the latency by nearly 50\%}, outperforming even the Bullshark protocol in partially synchronous settings. We observe this 50\% latency reduction even when increasing the number of nodes. This reduction is near-optimal because the vote-based BFT consensus needs at least $3$ rounds. Given that Tusk's average latency is $7$ rounds, and our scheme only commits on odd rounds, 50\% latency reduction is near optimal. 

Second, our algorithm reaches the computation power bottleneck earlier than other protocols. It happens at an estimated 64K input TPS. This is attributed to the increased metadata handling and transaction commitment on Board. In Figure~\ref{fig_latency_tps}, this bottleneck manifests as a latency increase at the rightmost point for the orange and red curves. Despite this, our additional load yields substantial gains in reduced latency and check of external validity, which Tusk and Bullshark will accomplish afterward. We predict that realistic network delay and node capability make this bottleneck more challenging. Considering that TPS does not affect our metrics, the following tests used a load of 200 TPS to reduce the computation load. 

The fast-commitment rate, defined as the ratio of fast-committed transactions to all transactions, is another significant metric for our scheme. Although not depicted in Figure~\ref{fig_latency_tps}, the \textbf{fast-commitment rate remains steady} at approximately 50\% of all transactions until reaching the computational bottleneck. This outcome aligns with our expectation that the fast-commitment rate should correlate solely with the network topology, \ie, the committee size and number of faulty nodes.

\begin{figure}[ht]
        \centerline{\includegraphics[width=\linewidth]{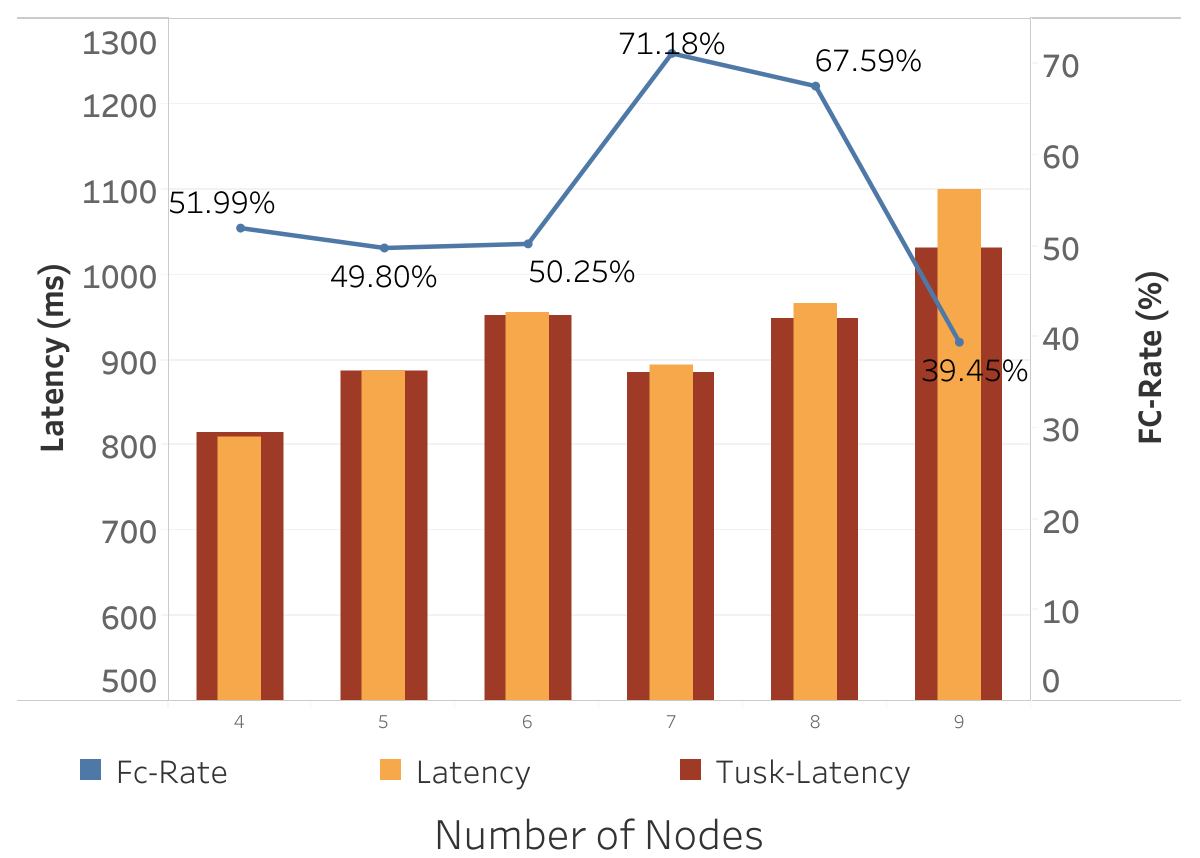}}
	\centering
	\caption{Formal latency and fast-committed rate changed according to the number of nodes.}
	\label{fig_scale_nodes}
\end{figure}

The scalability test in Figure~\ref{fig_scale_nodes} offers an alternative perspective by varying the number of nodes. Generally, the fast-commitment rate positively correlates with the maximum $f$ and slightly negatively correlates with $N$. When $N\in \{4,5,6\}$, we have $f=1$, and the fast-commitment rate is around 50\% of all transactions. While $N\in \{7,8,9\}$, we have $f=2$, which greatly improves the fast-commitment rate to a level of around 70\% of all transactions. This is because these additional nodes in the system may not contribute to leader selection but still help to increase the number of transaction votes.

\begin{figure}[ht]
	\centering
	\includegraphics[width=\linewidth]{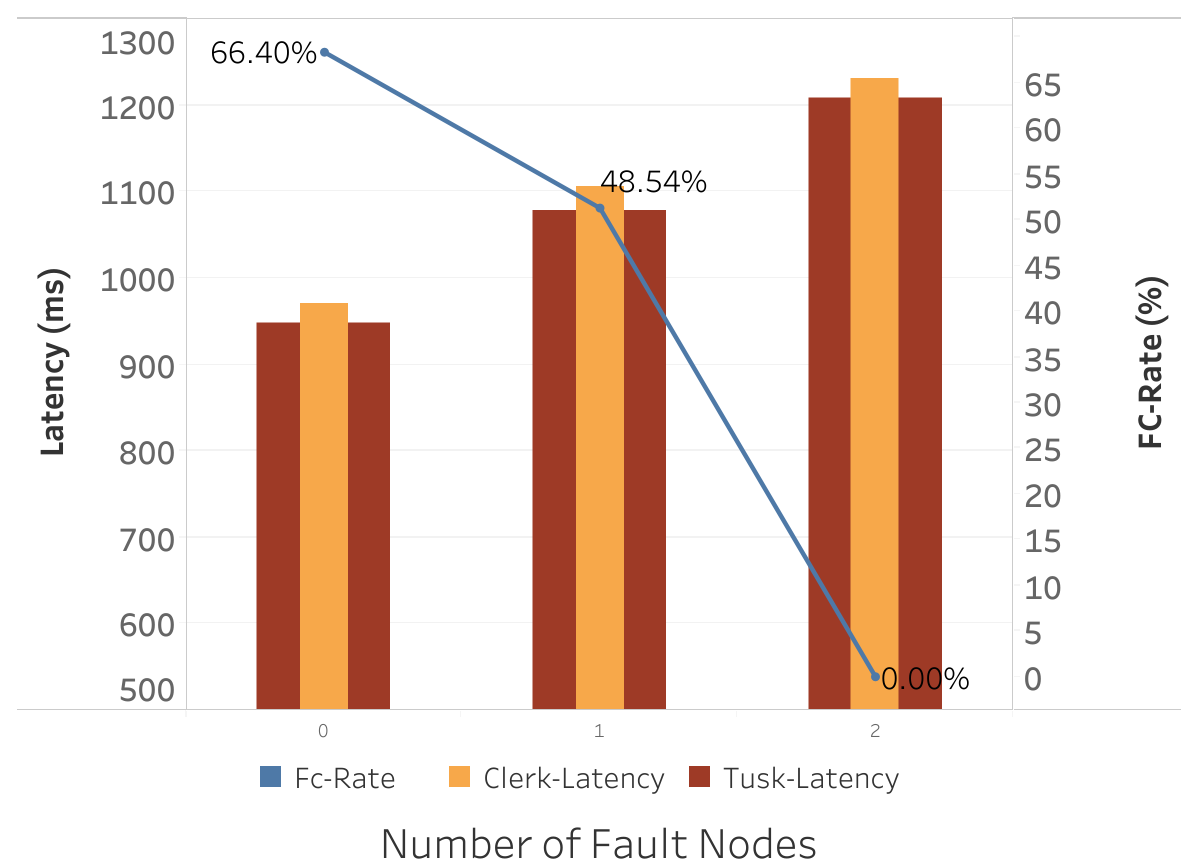}
	\caption{Formal latency and fast-committed rate changed according to the number of faulty nodes.}
	\label{fig_redundancy_gain}
\end{figure}

The faulty node test depicted in Figure~\ref{fig_redundancy_gain} provides the most compelling evidence. In this test, we gradually shut down nodes within an 8-node system. It was observed that the fast-commitment rate gradually decreased in response to node shutdowns. Ultimately, the rate reached zero, indicating that no transactions could employ the fast commitment channel when nodes reach consensus every two rounds. We omit simulation for Byzantine faulty nodes. This is because the reliable broadcast protocol in the Narwhal memory pool will not accept proposals from byzantine nodes. Therefore, byzantine nodes are the same as crashed nodes for Board and Clerk.

The scalability test shows that the latency of Clerk's formal commitment escalates more rapidly than that of the Tusk algorithm when scaling nodes. This outcome is anticipated, given that Board must manage an increased volume of votes and the corresponding counting process. It is also noteworthy that dummy transactions, which may be introduced due to concerns about Byzantine nodes, can further reduce the fast-commit latency as more proposals will include the transaction and increase the number of votes. Last, our Clerk implementation is a modification on Tusk for asynchronous settings. However, we can also use the partial synchronous setting and implement Clerk, based on Bullshark, reducing the latency of non-fast-committed transactions while keeping our fast-commitment channel advantages.

\section{Conclusion}\label{sec:conclusion}
    In our research, we introduce a new method for improving the latency of the Narwhal-based consensus algorithm. We propose the Board and Clerk Protocol as an efficient consensus algorithm for blockchain-aided systems that rely on the UTXO Model. Our algorithm's foundation is a novel commit rule, combined with a fast-commitment channel, which collectively can reduce the latency of a minimum of 50\% of transactions without incurring additional communication overhead. Our performed experimental results indicate that our protocol effectively leverages redundant nodes and transactions to minimize latency. Given that this redundancy originates from security requirements, the latency reduction essentially incurs without extra costs if the node computational power is not a limiting factor.

Furthermore, we outlined two strategies for achieving verifiable transaction outcomes without necessitating an increase in communication rounds. We regard this characteristic of our research as a crucial feature that a public blockchain system based on a consortium should support. 

Lastly, we hope our proposal may motivate researchers to explore the possibility of using the underlying data model to augment consensus algorithms akin to our UTXO-enhanced consensus algorithm.

\section*{Acknowledgment}
This work was sponsored by JST SPRING, Grant Number JPMJSP2108; JSPS KAKENHI, Grant Number JP23KJ0385 and JP21K11882; JST CREST, Grant Number JPMJCR2113.

\bibliographystyle{IEEEtranS}
\bibliography{bib/reference,bib/abbrev0,bib/crypto}
\end{document}